\documentstyle[aps,twocolumn,floats,epsf]{revtex}

\begin{document}

\wideabs{
\title{
 First-principles studies of kinetics in epitaxial growth 
 of III-V semiconductors
}

\author{
 P.~Kratzer, 
 E.~Penev, and 
 M.~Scheffler
}                     

\address{Fritz-Haber-Institut der Max-Planck-Gesellschaft, 
Faradayweg 4--6, D-14195 Berlin, Germany 
}

\maketitle

\begin{abstract}
We demonstrate how first-principles calculations using
density-functional theory (DFT) can be applied to gain insight into
the molecular processes that rule the physics of materials processing.
Specifically, we study the molecular beam epitaxy (MBE) of arsenic
compound semiconductors.  For homoepitaxy of GaAs on GaAs(001), a
growth model is presented that builds on results of DFT calculations
for molecular processes on the $\beta2$-reconstructed GaAs(001)
surface, including adsorption, desorption, surface diffusion and
nucleation.  Kinetic Monte Carlo simulations on the basis of the
calculated energetics enable us to model MBE growth of GaAs from beams
of Ga and As$_2$ in atomistic detail.  The simulations show that
island nucleation is controlled by the reaction of As$_2$ molecules
with Ga adatoms on the surface.  The analysis reveals that the scaling
laws of standard nucleation theory for the island density as a
function of growth temperature are not applicable to GaAs epitaxy.  We
also discuss heteroepitaxy of InAs on GaAs(001), and report
first-principles DFT calculations for In diffusion on the strained
GaAs substrate.  In particular we address the effect of
heteroepitaxial strain on the growth kinetics of coherently strained
InAs islands.  The strain field around an island is found to cause a
slowing-down of material transport from the substrate towards the
island and thus helps to achieve more homogeneous island sizes.
\end{abstract}
}

\section{Introduction}
\label{Introduction}

In the last decade, research in semiconductor heterostructures has
evolved into one of the most active fields in semiconductor physics.
Progress in this field owes much to advanced growth methods such as
molecular beam epitaxy (MBE) and metal-organic vapor phase epitaxy
(MOVPE).  Using these techniques, heterostructures from compound
semiconductor materials can be grown with high precision. The quantum
well and quantum dot structures obtained in this way are of
considerable technological interest, since they are used as active
components in devices, e.g. in high-electron-mobility transistors and
optoelectronics. The possibility to design well-defined nanostructures
or nanostructured materials by epitaxial growth techniques has raised
the desire to understand and eventually control growth on the atomic
scale. While important input to such an understanding comes from
probes with atomic resolution, in particular from scanning tunneling
microscopy (STM) studies, theoretical input from growth simulations is
equally important. Only a combined approach, using the results of
modeling to interpret experimental data, will enable us to bring to
light all the atomistic details involved in the fairly complex growth
process of a compound semiconductor material.
  
At the atomic level, understanding MBE and MOVPE requires knowledge
about the {\em chemical reactions} between the substrate surface and
the atoms or molecules used as reactants (frequently called
'precursors' in the MOVPE literature), in addition to knowledge about
the processes occuring {\em on} the surface, such as diffusion,
nucleation and attachment or detachment of adatoms to existing islands
and steps. In recent years, there has been considerable progress in
the understanding of surface chemical reactions using {\em ab initio}
methods.  In particular, total-energy calculations on the basis of
density-functional theory (DFT) have proven to be most helpful for
describing reactions in a fairly complex atomic environment, e.g. at a
reconstructed surface.  The dynamics of individual reactive encounters
with a surface have been modeled in great detail; either by
integrating the classical trajectory of a molecule over a few hundred
femtoseconds, or by the use of quantum-mechanical scattering
techniques \cite{GrSc98rev,Kroe98}. In contrast to elementary
reactions, when modeling epitaxial growth, the properties of interest
(e.g. the growth morphology) develop only over time scales of the
order of seconds and length scales of micrometers, while the ruling
microscopic processes operate in the length and time domains of 0.1--1 nm,
 and femto- to pico-seconds. Hence, the use of DFT calculations
for modeling atomistic aspects of growth has been hampered by the need
to bridge length and time scales by many orders of magnitude.
First-principles molecular dynamics studies, while being very useful
in the investigation of individual events on time scales shorter than
100 ps, are not suitable for accessing the time scales involved in
growth, nor can they tackle with the statistical interplay of the
numerous processes that are responsible for the outcome of a growth
experiment.  As will be shown in this paper, kinetic Monte Carlo (kMC)
simulations offer an efficient and accurate way to cope with this
difficulty.

The relevance of physical chemistry for the elementary steps in
compound semiconductor epitaxy has been realized more than twenty
years ago \cite{foxon:77}.  While deposition of elemental metal films
involves the diffusion and nucleation of a single species, MBE and, to
an even larger extent, MOVPE, involve reactions of molecular species,
either among each other or with the surface. Any of these elementary
steps could in principle be rate-limiting, or affect in some way the
growth morphology.  Principal considerations of this kind, exemplified
by kinetic Monte Carlo studies, led Ghaisas and Madhukar to the
distinction between diffusion-limited and reaction-limited growth
\cite{ghaisas:86,ghaisas:88} in the context of GaAs epitaxy.

However, most of the work in the last decade has put aside the
additional complexity introduced by surface reactions.  A single
'effective' species, usually assumed to be gallium, moving on a simple
cubic lattice \cite{shitara:92a,smilauer:93}, or sometimes on the more
physical face-centered cubic lattice \cite{grosse:00}, was employed
for modeling of GaAs epitaxy.  Mainly motivated by the different time
scales involved in the oscillations of the RHEED signal and in its
recovery after a growth interruption, a model with two species, Ga
atoms and GaAs molecules, was introduced
\cite{heyn:97,heyn:97a,heyn:99}.  Recently, Ishii {\em et al.} also
treated a two-component model with both Ga and As atomic species
\cite{ishii:98,ishii:99}. Both these two-component models restricted
themselves to a simple cubic lattice.  The aspect of surface
reconstruction was not addressed until recently when Itoh {\em et al.}
\cite{itoh:98} presented results of a more refined modeling including
the interactions between surface As dimers, as well as the anisotropy
of Ga diffusion on the reconstructed GaAs surface.
Adopting suitable model parameters guided by experimental input from a
detailed STM study, Itoh {\em et al.} were able to
simulate the morphology, the size and density of small islands in very
good agreement with the experimental observations.~\cite{itoh:00,itoh:01}  
Yet the
're-construction' of a growth model on an empirical basis, working
backwards from the analysis of experimentally observed growth
features, is a tedious and, given the complexity of the problem,
sometimes ambiguous procedure. Without knowledge of the most likely
microscopic processes, it is difficult to decide which of two
alternative elementary steps should be part of the growth model when
both give rise to similar growth features.  It appears much more
appealing to construct a growth model in a 'bottom-up' fashion,
starting from knowledge about the atomic processes obtained from
first-principles total-energy calculations.  As an additional benefit
from such an approach, we obtain an immediate identification between
the processes included in the model and the motion of atoms involved
in growth.  This is in contrast to empirical growth models, where
`effective' processes are used to rationalize experimental findings,
but difficulties may arise later when attributing these `effective'
processes to real physical processes.

In the epitaxy of metals, there has been considerable progress in
recent years triggered by a microscopic understanding of growth on the
basis of density-functional total-energy calculations
\cite{ruggerone:97,BoSt98,OvBo99,FiSc00}.  Although the homoepitaxy of
a metal on a close-packed surface appears at first to be a fairly
simple problem, detailed experiments in conjunction with kinetic
simulations on the basis of first-principles results have revealed a
number of fascinating details.  Despite the obvious fact that epitaxy
of compound semiconductors is an even more complex problem, both due
to their two-component nature and the complexity of their surface
structure, we feel encouraged by the progress due to the use of
first-principles methods achieved for metallic systems, and tempted to
extend this approach to molecular beam epitaxy of compound
semiconductors.

\section{Homoepitaxy of G\lowercase{a}A\lowercase{s}}\label{GaAs}
\subsection{Present understanding}
\label{prerequisites}

In this section, we shall describe a model for MBE homoepitaxy of GaAs
on the GaAs(001) substrate from Ga atoms and As$_2$ molecules which
has been constructed on the basis of density-functional total-energy
calculations.  We are aiming at a simulation of the atomistic
processes of island growth including all relevant microscopic details,
such as the surface reconstruction, different kinetic properties of
the species involved, etc., based on the understanding we obtain from
first-principles calculations of the energetics.  Specifically, we
model island growth on the well-known $(2 \times 4)$ reconstructions
of GaAs(001) that prevail under frequently used growth conditions.
Several theoretical investigations
\cite{chadi:87,qian:88,northrup:93,northrup:94,moll:96,schmidt:00}
have contributed to our present understanding of the atomic structure
of this surface.  The starting point for our modeling is the $\beta2(2
\times 4)$ reconstruction of GaAs(001). This moderately As-rich
surface reconstruction is terminated by pairs of As dimers, which
alternate with `trenches' running in the $[\bar 1 1 0]$ direction
(missing a pair of As dimers and two Ga atoms per unit cell in the two
topmost layers).  Recently, a combined theoretical and experimental
STM study has unequivocally shown that this reconstruction is the
ground state of the surface when carefully prepared at temperatures
around $550^{\circ}$C and conditions typical for MBE \cite{LaYa:99}.
Other $(2 \times 4)$ reconstructions that are energetically close to
$\beta2$ (see Ref. \cite{moll:96} and references given therein) and
could possibly show up on the surface during growth are the $\alpha$
and $\beta1$ (also just called $\beta$) reconstructions. They evolve
out of the $\beta2$ structure by adding two Ga atoms in the trench
($\alpha$) , and by further adding two As atoms, forming a third As
dimer in the top layer ($\beta1$).  Recently it has also been
discussed whether the $\alpha2$ structure, which results from the
$\beta2$ structure by desorbing one of the top layer As dimers, could
be observed under particular conditions
\cite{schmidt:00,RaBa:00,MiSr:00}.  Only when growth conditions very
different from the standard ones are employed, major changes of the
surface reconstruction leading to different symmetries can occur.  At
lower growth temperatures, as used e.g. for heteroepitaxy of InAs on
GaAs \cite{joyceASS:98}, one typically has a more As-rich chemical
environment in the growth chamber.  This gives rise to a $c(4 \times
4)$ reconstruction of the GaAs(001) surface, where the two topmost
layers consist entirely of As atoms \cite{biegelsen:90a}.  In the
other extreme, annealing of the GaAs(001) surface in vacuum results in
the Ga-rich $(4 \times 2)$ surface \cite{LeMo:00,KuMa:01}.
Homoepitaxy on theses surfaces, as well as the phase transitions
between structures of different symmetry, is beyond the scope of this
paper.

Both experimental knowledge and previous theoretical studies agree
that arsenic and gallium behave quite differently on the GaAs(001) $(2
\times 4)$ surface.  While Ga atoms adsorb with unit sticking
probability, As$_2$ molecules only stick to the surface after Ga has
been deposited \cite{foxon:77,tok:97}.  These findings have been
substantiated by recent calculations using density-functional theory
\cite{KrMo:99,MoKr:99}.  While Ga adatoms were found to bind strongly
(between 1.5 and 2.1 eV at various surface sites) \cite{KrMo:98}, the
binding sites for As$_2$ molecules at a perfect GaAs(001)-$\beta2$
surface are either much weaker, or only accessible after traversing a
considerable activation barrier \cite{MoKr:99,MoKr:02}.  The rather
weakly bound states of As$_2$ present on an ideal GaAs(001) surface
are not sufficiently long-lived at standard growth temperatures to
give rise to incorporation of molecules, but lead to temporary
trapping of molecules close to the surface.  However, as soon as Ga
adatoms are present on the surface, stronger binding sites for the
As$_2$ molecules arise that are accessible from the gas phase without
a barrier \cite{MoKr:99}.
On the Ga-precovered GaAs(001)-$\beta2$ surface, it is not compulsory
for the As$_2$ to dissociate upon adsorption, because a gas-phase
As$_2$ molecule can attach itself to the Ga adatoms and get
incorporated by transforming itself into a surface As dimer.  The
growth model presented here is based on this low-energy pathway to
arsenic incorporation.  The weaker binding sites for As$_2$ on the
pristine $\beta2$ surface are accounted for by introducing a
short-lived molecular precursor state.\footnote{The term 'precursor'
is used here in the sense as in surface science, meaning a weakly
bound molecular state preceeding the eventual transition to a
chemisorbed (often dissociated) state.}  The As$_2$ molecules trapped
in this state have the chance to explore an average number of surface
sites, before they either desorb or find excess Ga to attach to.  We
account for this effect of the precursor in our model by working with
an effective As$_2$ flux that is a factor of about 100 higher than the
external flux from the As$_2$ source.  The role of the As$_2$
precursor in growth is in accordance with previous experimental
findings \cite{foxon:77} as well as with computer simulations \cite
{itoh:98}.  We also point out that the GaAs(001) surface, given that
the $\beta2$ reconstruction with moderate additional Ga coverage
persists during growth, offers only a very limited number of strong
binding sites for As$_2$ molecules.  For sufficient As incorporation
to occur during growth, it is plausible that a mobile precursor state
assists the As$_2$ molecules in finding the suitable sites with excess
Ga needed for their adsorption.  In accordance with previous
investigations~\cite{itoh:98}, we further assume that the mobility of
arsenic on the surface is mainly {\em not} due to diffusion of As
adatoms, but originates from As$_2$ in the mobile {\em molecular}
precursor state.  In fact, single As adatoms are not included in the
present modeling, because their formation would require breaking of
the strong As--As molecular bond (4 eV), and hence is an unlikely
event.

For the Ga species, on the other hand, the mobility is completely
determined by surface diffusion.  Hence, a detailed account of Ga
diffusion is a prerequisite for modeling the morphology of the growing
surface.  Ga mobility is essential for growth because sites for strong
As$_2$ chemisorption are created only after a suitable local
arrangement of Ga adatoms.  In the context of DFT calculations, the
potential energy surface (PES) of an adatom is the adequate concept to
address the issue of diffusion. DFT calculations of potential energy
surfaces for various GaAs(001) surfaces are available from the
literature, both for Ga diffusion
\cite{shiraishi:96,kley:97,Lepage:98} as well as for As diffusion
\cite{seino:99}.  Most importantly, it was found (see
Ref. \cite{kley:97}) that Ga has its strongest adsorption site inside
As surface dimers, thereby breaking the dimer bond.  The picture
evolving from these computational results suggests that a possible
role of the Ga atoms in growth could be the breaking-up of existing As
surface dimers, and the formation of locally Ga-rich environments that
act as selective sites for adsorption of As$_2$ molecules from the gas
phase.

In the following section, we investigate the consequences of the
energetics, as obtained from DFT calculations, for the statistical
interplay of the various surface species during growth. In particular,
we are interested in the statistics of island nucleation and its
dependence on the growth conditions.  To this end, we have performed
kinetic Monte Carlo simulations for an atomistic growth model that
includes the most relevant microscopic details.

\subsection{Kinetic Monte Carlo simulations}
\label{kMC}

In order to simulate epitaxial growth, we employ a mathematical model
of the crystal, describing it as a three-dimensional lattice with the
zincblende crystal structure.  Ga and As atoms are occupying different
sublattices.  It is required that the position of each atom in the
simulation can be associated with a lattice site.
The crystallographic directions $[110]$, $[\bar 1 1 0]$ and $[001]$ 
define the $x,y$ and $z$ axis of the coordinate system.  
The crystal grows in the $z$ direction.  With
each lattice site we associate a discrete variable describing the
state of this site.  In the present case, a site can be either empty,
occupied by a Ga, or an As atom.  This discrete variable also carries
the information about the local bonding of surface atoms different
from those of bulk atoms due to surface reconstructions.  Arsenic
atoms in the topmost anion layer form As dimers, unless a Ga atom is
sitting in them.  The starting surface is prepared in the
thermodynamic ground state, the $\beta2$ reconstruction of GaAs(001).
Some randomly selected unit cells, however, are given in the $\beta1$
reconstruction, as this is energetically very close to the $\beta2$
reconstruction.  The probability for a surface unit cell to be in the
$\beta1$ structure when the simulation is started is determined from a
Boltzmann distribution, where the calculated difference in surface
energy per unit cell between the $\beta1$ and $\beta2$ reconstruction
(360 meV \cite{moll:96}) is used in the exponent.

\begin{figure}[tb]
\epsfxsize=0.9\columnwidth
\centering
\epsffile{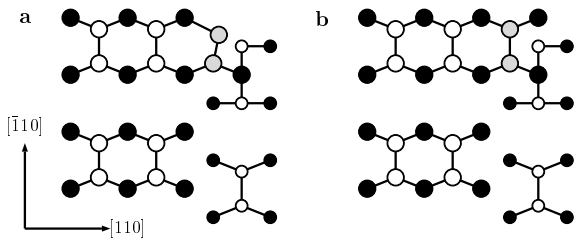}
\vskip0.5cm
\caption{ (a) As$_2$ molecule (grey) adsorbed on a single Ga adatom
that has split a trench As dimer, establishing three As--Ga backbonds,
thereby forming a Ga-As-As-Ga$_2$ complex; (b) adsorbed on two Ga
adatoms in the trench, establishing four As--Ga backbonds, thereby
forming a Ga$_2$-As-As-Ga$_2$ complex. Arsenic atoms of the $\beta2$
reconstruction are shown as empty circles, gallium atoms as filled
circles.}
\label{Asdes}
\end{figure}

Time evolution proceeds by discontinuous changes of the occupation of
discrete lattice sites. These events may be either adsorption or
desorption of atoms or molecules, or the hopping of an atom from one
site to another. Each event is characterized by a rate, which is
determined prior to starting the simulation.  We use a parameter-free
approach to determine the rates on the basis of the energetics
obtained from DFT calculations.
In the following we describe the events included in the present
simulation in more detail.
\begin{itemize}
\item Ga adsorption

Ga adsorption occurs with a probability given by the Ga flux times the
area of the sample.  A Ga flux of 0.1~ML/s typical for MBE growth is
used in the simulations.  A random site in the $xy$-plane is selected
for Ga adsorption; the layer where the adsorption takes place is
determined by the local height of the surface. If this site is
occupied or unsuitable for adsorption, one of the four nearest
neighbor sites is selected. If all of these are occupied or
unsuitable, the Ga atom is assumed to be reflected into the gas phase.
Other Ga desorption events are not considered in the model.
   
\item Ga surface diffusion

Investigation of the Ga adatom diffusion starts with the calculation
of the potential-energy surface (PES), which requires the
determination of the optimum geometry and the total energy of the
adatom and the substrate for a number of fixed lateral positions of
the adatom.  For each position, a DFT calculation including a
structural optimization is performed.  Special care must be taken of
the fact that the PES may in general be a multi-valued function of the
adatom position, in particular if the adsorbate can occupy
substitutional or subsurface sites, or if it induces bond breaking in
the substrate. For an example, we refer the reader to
Ref.~\cite{kley:97}.  The energy minima in the PES obtained in this
way are mapped onto sites of the lattice used for the kMC simulation.
The rates for hops between two lattice sites are determined from the
energy difference $\Delta E$ between the minimum corresponding to the
initial position of the adatom and the relevant saddle point for
hopping to a neighboring minimum.  Using transition state theory, it
is also possible to determine the prefactors from first principles
(see, e.g. Ref.~\cite{ratsch:96}).  For Ga diffusion on GaAs(001),
previous calculations have found prefactors of the order of 10$^{12}$
to 10$^{13}$ s$^{-1}$ for all hopping processes considered
\cite{kley:97,kley:98}.  Thus, the effect of possible prefactor
variations is expected to be small. In the present work, we use a
common prefactor of $10^{13}$ s$^{-1}$ for all processes.
Due to the surface reconstruction, the rates for hopping of Ga adatoms
are highly site-specific. We include 10 different hopping processes
for an isolated Ga adatom in our model.  For Ga atoms with one or two
Ga next-nearest neighbors, different hopping rates modified by the
binding energy between Ga atoms in these configurations are used
\cite{KrMo:99}.  In total, we differentiate between 29 hopping
processes of Ga.

\item Ga incorporation

The DFT calculations show that binding of a Ga atom in some
configurations is so strong that it is unlikely that the Ga atom will
leave this site again during the time span of the simulation. Such an
incorporation event takes place
\begin{itemize}
\item if As is adsorbed above the Ga atom,  
\item if it forms a dimer with a Ga neighbor in [110] direction in the
top layer,
\item if it sits in the interior of a string of Ga atoms in $[\bar 1 1
0]$ direction
\end{itemize}

\item As$_2$ adsorption

In the simulation, adsorption of As$_2$ can occur only at specific
surface sites.  At these sites, the adsorbing As$_2$ molecule becomes
incorporated as an As surface dimer, with the dimer axis oriented
along [$\bar 1 1 0 $].  This is motivated by DFT calculations which
show that the binding is sufficiently strong only if the As$_2$ forms
three or four backbonds to the surface.  Figure~\ref{Asdes}
illustrates two examples of sites where an As$_2$ molecule (shown in
grey) has adsorbed.  The As dimer has filled a pair of empty sites in
the anion sublattice, surrounded by at least three (out of possible
four) Ga atoms on the cation sublattice.  Upon adsorption, a dangling
bond of each Ga atom is converted into an As--Ga bond.  Possible
As$_2$ adsorption at other, weaker binding sites are considered as
part of the As$_2$ precursor state, and are accounted for by a
renormalized effective As$_2$ flux in our model.  Since there are Ga
atoms with dangling bonds in both sidewalls of the trench, a single
additional Ga adatom in the trench is sufficient to form an
arrangement with three Ga dangling bonds, and hence to create an
adsorption site for As$_2$.  This is in contrast to As$_2$ adsorption
in the top layer, where at least three adatoms in adjacent sites are
needed for that purpose.  Thus As$_2$ adsorption in the trench is more
likely to occur in the early stages of growth.  However, adsorption of
two adjacent As$_2$ molecules parallel to each other in the trench is
disallowed in our model, unless at least one Ga atom has already
adsorbed next to the trench. This is motivated by the results of a DFT
calculation~\cite{KrMo:99} showing that adsorption of a second As
dimer next to an existing one in the trench (which would level out the
trench) is unfavorable.  Previous DFT calculations \cite{shiraishi:98}
demonstrated that the presence of Ga atoms in the neighborhood of the
prospective adsorption site for As$_2$ significantly enhances the
binding energy of the As$_2$ at this site.  Therefore, adsorption of
the last missing As dimer that levels out the trench is enabled in the
simulations only after a Ga atom has adsorbed in its vicinity.

\item As$_2$ desorption

Our model differentiates between three As$_2$ desorption events with
different rates, depending on the chemical environment the As$_2$ is
bonded to. These are
\begin{itemize}
\item desorption of As$_2$ from sites where it had three backbonds to
Ga atoms ($\Delta E=1.9$~eV).
\item desorption of As$_2$ from sites where it had four backbonds to
Ga atoms, but was misaligned with an adjacent As dimer row ($\Delta
E=1.9$~eV).
\item desorption of one of the outermost As$_2$ from a row of three
adjacent As surface dimers (local $\beta$ structure, $\Delta
E=2.4$~eV).  In this case, the desorbing As dimer had four backbonds
to Ga.  Despite the high barrier for this process, it is necessary to
include it to make it possible for the $\beta1$ structure to decay to
the ground state of the surface, which is the $\beta2$ structure.
\end{itemize}
In all cases an Arrhenius law for the desorption rate with a prefactor of 
$10^{13}$ s$^{-1}$ is used, following the analysis 
of experimental desorption data in the literature \cite{BaCr:92}.  
\end{itemize}

\subsection{Results of the simulations}
\label{simulations}

The simulations reveal that island growth is governed by the interplay
of processes that occur on a hierarchy of time scales.  In the
following, we illustrate the situation for deposition at a surface
temperature of 800~K.  The shortest time scales ($10^{-12}$~s --
$10^{-9}$~s) are set by the hopping diffusion of Ga adatoms.  Since
hopping in $[\bar 1 10]$ direction along the trenches is faster than
hopping from one trench to another in [110] direction, Ga diffusion on
the GaAs(001)-$\beta2(2 \times 4)$ surface is anisotropic.  The
terminating As dimers of the $\beta2$ reconstruction, in particular
those in the trenches, can act as traps for diffusing Ga atoms, where
they are immobilized for $10^{-8}$~s.  At sites where a Ga atom has
been adsorbed in a trench site, a gas-phase As$_2$ molecule can adsorb
readily by establishing three bonds to surface Ga atoms, one of them
to the Ga adatom and two other bonds with unsaturated Ga atoms that
are already present in the $\beta2$ reconstruction in the slopes of
the trenches.  We refer to the structures resulting after As$_2$
adsorption as Ga-As-As-Ga$_{2}$ complexes (Fig. \ref{Asdes}a).  At
temperatures below 800~K, such complexes have a mean lifetime of 0.1~s
or more, long enough to react with another diffusing Ga adatom.  This
transforms them into even more long-lived Ga$_{2}$-As-As-Ga$_{2}$
complexes (Fig. \ref{Asdes}b), that can be considered as unit cells in
a local $\beta1$ reconstruction.  The meta-stability of these
intermediate structures is related to the local fulfillment of the
electron counting rule, i.e. no induced electronic states in the
principal band gap are present after an atomic arrangement
corresponding to local $\beta1$ reconstruction has been reached.  We
note that the importance of intermediate structures fulfilling the
electron counting rule was pointed out earlier \cite{ito:96,ito:97a}.

\begin{figure}[tb]
\epsfxsize=0.8\columnwidth
\centering
\epsffile{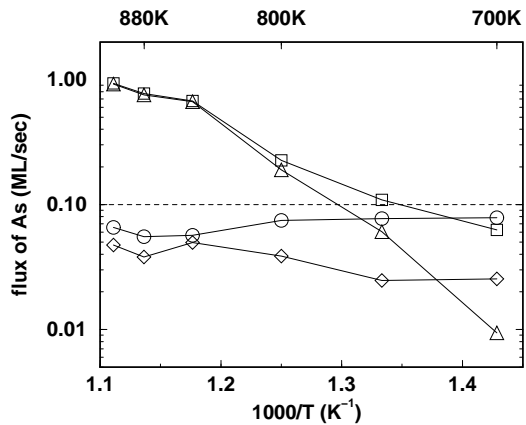}
\vskip0.5cm
\caption{Arsenic flux as a function of temperature during stationary
island growth.  Net As incorporation (circles) results from a balance
between desorption (triangles), and adsorption forming three Ga-As
backbonds (squares), or four Ga-As backbonds (diamonds).  The maximum
possible growth rate determined by the Ga flux is indicated by the
dashed horizontal line.}
\label{Asflux}
\end{figure}

The role of the As$_2$-Ga complexes in growth becomes apparent from an
analysis of the balance between adsorbing and desorbing As$_2$ fluxes
during a kMC simulation.  The speed of incorporation of arsenic into
the substrate is given by the difference between the adsorbing flux
and the desorbing flux $F_{\rm des}$.  In Fig.~\ref{Asflux}, we
decompose the adsorbing flux into a contribution $F_{\rm ads}^{(3)}$
from Ga-As-As-Ga$_{2}$ and $F_{\rm ads}^{(4)}$ from
Ga$_{2}$-As-As-Ga$_{2}$ complexes, respectively. These are compared to
the net incorporated flux, $F_{\rm As} = F_{\rm ads}^{(3)} + F_{\rm
ads}^{(4)} - F_{\rm des}$.  We find that, at $T=$~800~K, the main
route for chemisorption of incoming As$_2$ molecules is the formation
of Ga-As-As-Ga$_{2}$ complexes.  These are formed by As$_2$ reacting
with single Ga adatoms at trench sites.  The simultaneous reaction of
As$_2$ with two adjacent Ga adatoms, leading to the
Ga$_{2}$-As-As-Ga$_{2}$ complex, takes place as well, but contributes
less to the net incorporation.  The Ga-As-As-Ga$_{2}$ complexes must
be considered as transient species that are only formed during
growth. They either decay by desorption of As$_2$, or become
stabilized by addition of another Ga adatom.  In the later case, we
say that the chemisorbed As$_2$ has become incorporated into the
surface, because it has achieved four backbonds to Ga, and desorption
from this state is an infrequent event.  Since this incorporated state
is typically reached via the intermediate Ga-As-As-Ga$_{2}$ complex,
the latter plays the role of a `doorway' state for the incorporation
 of As$_2$.
Further support for its importance for As$_2$ incorporation comes from
the observation that the sticking probability of As$_2$ molecules for
this pathway is first order in the Ga coverage, because a single Ga
atom in a trench site is sufficient to initiate the formation of the
Ga-As-As-Ga$_{2}$ complex.  This is in agreement with the
experimentally observed reaction order \cite{foxon:77}.
Figure~\ref{Asflux} shows that the net incorporation rate of arsenic
decreases slightly at $T> 800$~K, and the adsorption due to
Ga-As-As-Ga$_{2}$ complexes is almost completely consumed by a
desorption flux of the same magnitude. At $T=850$~K, the net arsenic
incorporation is mainly due to direct reactive formation of the more
strongly bound Ga$_{2}$-As-As-Ga$_{2}$ complexes.  Eventually, at the
highest temperatures, even these structures, forming part of groups of
three As surface dimers (local $\beta1$ reconstruction), become
unstable against desorption.  At $T=850$~K, the rate for this
desorption process ($\Delta E=2.4$ eV) is 0.06 s$^{-1}$.  This
compares favorably with reported desorption processes associated with
the formation of the $\beta2$ surface structure in this temperature
range \cite{PrTr:00}.
We note that the chemisorbed flux of arsenic, $F_{\rm ads}^{(3)} +
F_{\rm ads}^{(4)}$ over the whole temperature range in
Fig.~\ref{Asflux} lies between 0.1~ML/s and 1~ML/s. For a Ga flux of
0.1~ML/s, this corresponds to a V/III ratio between 1 and 10, in
agreement with frequently used experimental conditions for MBE growth.
The much higher effective As$_2$ flux of 100~ML/s used in the
simulation just reflects the average number of attempts that an As$_2$
molecule in the mobile precursor state makes to find a suitable site
for chemisorption before it desorbs from the precursor.  Taking the As
flux that reaches the chemisorbed state as a measure, the impinging
flux used in the simulation leads to values for arsenic chemisorption
in the right order of magnitude.
 
\begin{figure}[tb]
\epsfxsize=0.8\columnwidth
\centering
\epsffile{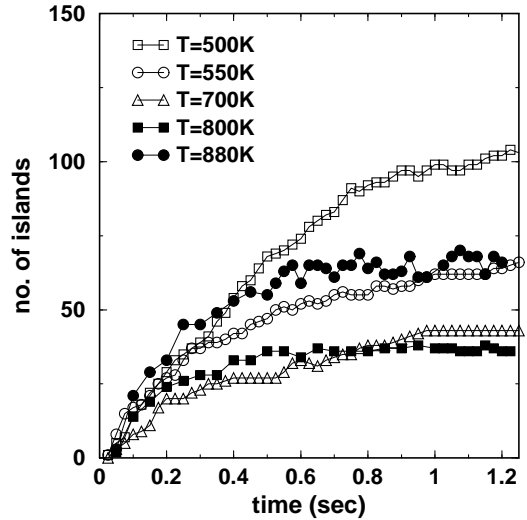}
\vskip0.5cm
\caption{Time evolution of the island densities in the new layer in a
simulation area of $160 \times 320$ surface lattice constants at
various temperatures
for a Ga flux of 0.1~ML/s and an As$_2$ pressure of $8.5 \times
10^{-7}$ bar.}
\label{islevol}
\end{figure}

For a further test of the growth model, we compare the island
densities in the simulations with measured island densities.  The
comparison is made for the saturated island density reached after some
fraction of a monolayer has been deposited.  When this island density
is reached, attachment of a deposited Ga atom to an existing island
typically takes place before the adatom could take part in a new
nucleation event.  This leads to a very small rate of nucleation that
is counterbalanced by a simultaneous decrease in the number of islands
due to coalescence.  Depending on temperature, the saturation occurs
in our simulations after one second or more, corresponding to
deposition of about 0.1~ML (see Fig.~\ref{islevol}).
We note that some caution is needed when relating the temperature
scale in the simulations to measured temperatures.  While DFT
calculation can reliably predict the relative sequence of barrier
heights, absolute barrier heights are not available yet with an
accuracy better than $\pm 0.1$~eV.  A 10\% error in all barriers, for
instance, would shift the temperature scale by the same relative
amount.  Because of this uncertainty, we decided to perform
simulations over a wide temperature range, from 500~K up to 900~K.  In
addition, systematic studies of the island density as a function of
substrate temperature are of interest because they have been widely
used to derive the diffusion constant for adatoms as well as its
temperature dependence from measurements of the island density
\cite{Brun98,MoKl91,MoKl92,LaBu:00}.  The extraction of the diffusion
constant from the experimental data is made under the assumption that
nucleation theory is valid in its variant describing diffusion-limited
attachment.  In this case, the island density is a monotonically
decreasing function of the growth temperature.

\begin{figure}[tb]
\epsfxsize=0.8\columnwidth
\centering
\epsffile{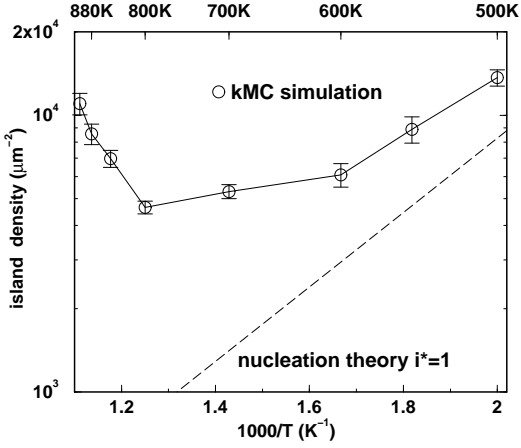}
\vskip0.5cm
\caption{Saturation island density as a function of the inverse growth
temperature.  The dashed line shows the prediction of nucleation
theory for diffusion-limited attachment and a critical nucleus size
equal one, using the Ga diffusivity along $[110]$, $D=D_0
\exp(-0.8{\rm eV}/k_{_{\rm{}B}}T)$.  The fluxes of Ga and As$_2$ are
the same as in Fig.~\protect\ref{islevol}.}
\label{isldens}
\end{figure}

In order to check whether island nucleation on GaAs can be described
by this frequently used theory, results from a systematic study of the
island density as a function of growth temperature are displayed in
Fig.~\ref{isldens}.  We find a decrease of the saturation island
density when the substrate temperature is increased from 500~K to
600~K.  This behavior is consistent with nucleation theory, if one
assumes diffusion-limited nucleation of islands with a critical
nucleus size $i^* = 1$.  However, in the simulations the island
density comes out to be almost constant above 600~K, and even
increases again for $T > 800$~K.  The reason for this unusual
non-monotonic behavior becomes clear when one considers the balance of
material flow for As at the growing surface shown in
Fig.~\ref{Asflux}.  At low temperatures and sufficiently high As$_2$
partial pressures, supply is dominated by $F_{\rm ads}^{(3)}$, i.e. by
the complexes of three Ga+As$_2$.  At higher temperature, the
initially formed Ga-As-As-Ga$_{2}$ complexes become unstable against
As$_2$ desorption. A decreasing fraction of them can stabilize by
capturing a Ga adatom and forming a more long-lived
Ga$_{2}$-As-As-Ga$_{2}$ complex.  Due to the decay of the
Ga-As-As-Ga$_{2}$ intermediate at high temperatures, island edges are
effectively less `sticky' for diffusing Ga adatoms than at low
temperatures.  This, in turn, leads to a higher density of mobile Ga
adatoms, a higher nucleation rate of new islands, and thus to the
observed rise of the saturation island density at temperatures above
800~K.

\begin{figure}[tb]
\epsfxsize=\columnwidth
\epsffile{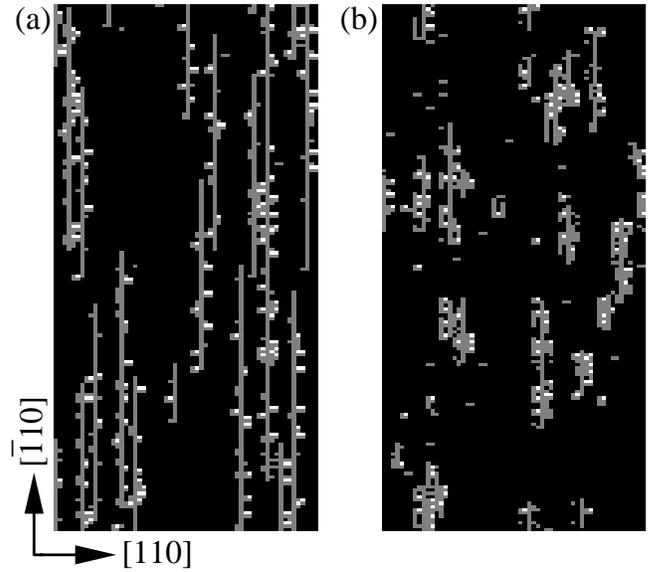}
\vskip0.5cm
\caption{Surface morphology shown on a scale of $80 \times 160$
lattice sites after growth of 0.1~ML GaAs at $T=700$~K (a) and
$T=900$~K (b). Islands are shown in grey, As dimers deposited in the
new layer in white.}
\label{morph}
\end{figure}

Additional information can be obtained from inspecting how the island
morphology evolves during the simulations.  Fig.~\ref{morph}a shows
that growth below $T=700$~K proceeds mainly by trench-filling. At
higher temperatures, an increasing fraction of islands also extend
into the regions inbetween trenches, forming a new layer of material.
The increasing amount of material grown in this new layer is related
to the reversibility of As$_2$ adsorption and desorption from
Ga-As-As-Ga$_{2}$ complexes at temperatures above $T=800$~K.  As a
consequence, structures grown in the trenches can dissolve again, and
the material remobilized in this way gets attached to islands that
extend into the new layer.  The influence of growth temperature
becomes apparent by comparing the island morphology at $T=700$~K in
Fig.~\ref{morph}a and $T=900$~K in Fig.~\ref{morph}b.  In
Fig.~\ref{morph}a most of the deposited material fills in the trenches
of the $\beta2$ reconstructions (grey vertical columns), while a
smaller fraction appears in the new layer.  In Fig.~\ref{morph}b, we
see filled trenches only if they are part of islands that extend into
the new layer.  The latter is consistent with STM images recorded
after growth of $~0.1$~ML \cite{bell:99c}.
The simulated island densities agree with those obtained from STM
experiments.  Experimentally, the island density increases with higher
V/III ratio, from 3400 to 6600~$\mu$m$^{-2}$ \cite{bell:99c,joyce:99}
for growth with an As$_2$ beam.  The simulations yield an island
density of 4600~$\mu$m$^{-2}$ in the temperature range between $T =
700$ and 800~K, where the temperature dependence is weak.  Reducing
the As$_2$ flux by an order of magnitude at 700 K in the simulations
resulted in a decrease of the island density, in accordance with the
experimental findings.

Measurements of the island density after sub-mono\-layer deposition
have been used to derive an experimental estimate of the diffusivity
of adatoms \cite{MoKl91,MoKl92,LaBu:00}.  In the data analysis, it is
usually assumed that the island density is described by nucleation
theory in its particular form applicable to diffusion-limited
nucleation with a critical nucleus $i^*=1$.  In this case, the island
density as a function of temperature $T$ is predicted to increase like
$\left(F/D(T)\right)^{1/3}$, where $F$ is the flux of Ga atoms, and
$D(T)$ is the diffusivity.  We note that the island density at typical
growth temperatures both in experiment and in our simulations is an
order of magnitude higher than predicted by these conventional
arguments.  In contrast to this prediction, but in agreement with our
simulations, a homoepitaxial film of GaAs evolves by coalescence of an
unusually large number of small islands elongated along the $[\bar 1 1
0]$ direction.  Our investigation thus demonstrates that some caution
is in order when deriving information about diffusivities from
measured island densities.  Given the rather complex $\beta2(2 \times
4)$ reconstruction of the GaAs(001) surface, one would expect that
coalescence of islands gives rise to disorder in the film when two
islands whose $(2 \times 4)$ unit meshes are not aligned start to
touch each other.  In order to achieve good film quality, it is
therefore desirable to keep the nucleation density of islands as low
as possible. Our simulations indicate a minimum of the island density
at a growth temperature of about 800 K, where desorption of As$_2$
starts to become relevant on the time scale of deposition.  The
temperature window frequently used by MBE crystal growers is a
compromise between the Ga adatom mobility and the lifetime of
As$_2$--Ga complexes that yields the lowest possible island density.

Summarizing the Section about homoepitaxy of GaAs, we stress that our
main goal is an understanding of the consequences of DFT results for
the growth kinetics of GaAs at various temperatures.  In contrast, the
atomistic growth model of Itoh and co-workers (Ref.~\cite{itoh:98},
for a detailed description of this model see Ref.~\cite{itoh:01}) is
based on parameters that are obtained by adjusting experimental and
simulated island morphologies at a single growth temperature, 873 K.
The information available from the DFT calculations motivated us to
include several features in our simulations that were not considered
previously: Most importantly, the incorporation of As$_2$ via an
intermediate state where the arsenic molecule forms three bonds to the
surface (rather than four) has not been considered before. Moreover,
we include a detailed description of the Ga diffusion processes in our
modelling.  In our simulations, formation of new structures is
possible by agglomeration of Ga atoms and As$_2$ molecules both in the
trenches and in the new layer; and the growth conditions decide about
the relative importance of these two possibilities.  This is in
contrast to earlier simulations guided by experimental results
\cite{itoh:98}, which took the experimentally observed island
morphologies as evidence for a nucleation process starting exclusively
in the new layer.


\section{Heteroepitaxy of I\lowercase{n}A\lowercase{s} on G\lowercase{a}A\lowercase{s}}
\label{InAs}

From the technological point of view, epitaxial growth is of
considerable importance because it has made it possible to grow
semiconductor heterostructures in a controlled way.  For instance,
heterostructures of III-V compounds have been used to achieve planar
confinement of charge carriers in quantum wells and to realize a
two-dimensional electron gas with high carrier mobility.  In recent
years, the heteroepitaxial growth of InAs on GaAs has received much
attention in the context of fabrication of quantum dots. By depositing
a few monolayers of InAs on GaAs(001), spontaneous evolution of
nanometer-sized islands (so-called self-assembly) was observed. After
growing a capping layer of GaAs on top the nanostructured film, the
buried islands, consisting mainly of InAs, can be used for confinement
of charge carriers in all three spatial dimensions, due to the lower
band gap of InAs compared to the surrounding GaAs matrix.  Under
judiciously chosen experimental conditions, the deposited material
forms elastically strained, defect-free three-di\-men\-sional islands
(for a review, see e.g. Ref.~\cite{shchukin:99}).  After overgrowth
with a capping layer of suitable material, and possibly after
repeating the procedure of islanding and overgrowth several times, the
resulting quantum dot (QD) structures lend themselves to a variety of
applications, foremost in optoelectronic devices such as
light-emitting diodes and quantum dot lasers \cite{kirstaedter:94}.
The latter have been based on stacked layers of QDs separated by thin
smooth capping layers. For the use of these nanostructures in laser
devices, it is crucial that the islands resulting from the growth
process are homogeneous in size because a spread in the size
distribution would result in non-resonant optical transitions in the
individual QDs, and thus spoil the basic operating mechanism of the
laser.

To a first approximation, the self-assembly of islands can be
understood as a form of growth in the Stranski-Krastanow regime.
According to this point of view, based on thermodynamic arguments, the
deposited material starts to grow as a homogeneous wetting layer,
because it has a lower surface energy than the substrate.  Only after
a critical thickness of this layer has been exceeded, the growth of
three-dimensional islands that allow for partial elastic strain
relaxation becomes energetically preferable to the two-dimensional
film growth. These three-dimen\-sio\-nal islands, once nucleated,
start to grow on the expense of the wetting layer, thereby competing
for the available material, until they have reached an optimum size
\cite{wang:99,wang:00a}.  This description of three-dimensional island
growth is rather general since it employs mainly thermodynamic
arguments.  Kinetics enters in this theory only as an external
parameter, the island density, that is determined during the
nucleation phase and assumed to remain constant during
three-dimensional growth of the islands.  A more detailed picture of
the formation process should not only account for nucleation kinetics,
but also for the kinetic aspects of growth.  While the stochastic
nature of the nucleation process is expected to give rise to some
irregularities in the spatial arrangement as well as to a broad size
distribution of the islands, it has been demonstrated experimentally
that the heteroepitaxial self-assembly process leads to more regular
structures, sometimes with a narrow size distribution.  In particular,
improved homogeneity is found in stacked QD systems when going from
the seed layer to higher and higher layers in the stack.  Since these
features are essential for the usefulness of the self-assembled
nanostructures as quantum dots and for their envisaged application in
future electronic devices, it is tempting to speculate that they
constitute intrinsic features of the growth process.  Various
mechanisms that could lead to narrow size distributions have been
suggested in the literature
\cite{shchukin:95,priester:95,chen:96,jesson:98,wang:99,wang:00a}.

Here, we shall discuss the role of strain for the evolution of
islands.  Each coherently strained island introduces a strain field
around it in the substrate, leading to lateral elastic interaction
between uncovered islands when a single sheet of quantum dots is
grown.  In addition, in multi-sheet arrays of quantum dots, strain
gives rise to interactions between islands in different sheets
separated by the capping layer.  We will come back to this issue
later.

\begin{figure}[tb]
\epsfxsize=\columnwidth
\epsffile{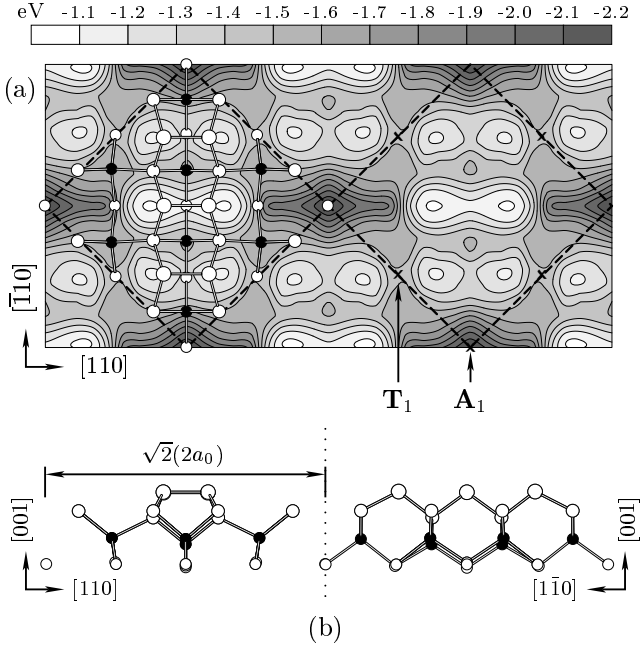}
\vskip0.5cm
\caption{(a) Potential-energy surface for an In adatom on the
GaAs(001)-$c(4\times4)$ surface. The positions of the atoms in the
four topmost layers in the clean surface unit cell are indicated (As:
empty circles; Ga: filled circles). The dashed boxes indicate the
surface unit cell.  Side views are shown in panel (b).}
\label{InGaAsPES}
\end{figure}

For uncovered islands in a single sheet, earlier studies in the
literature have found that the direct elastic interaction of islands
becomes important only at rather high island densities, when the
distance between islands is comparable to their base length
\cite{shchukin:98,ponchet:98}.
However, in addition to this direct interaction, strain also affects
the adatom density on the substrate surface during growth.  For
adatoms that are {\em locally} in equilibrium with a wetting layer on
the substrate, a thermodynamic treatment within a mechanical continuum
model, introducing a spatially varying chemical potential for the
adatoms, has been proposed (see,
e.g. Refs.~\cite{srolovitz:89,xie:95}).  Yet, little is known about
the effect of strain on the {\em kinetics} of growth due to changes of
the diffusivity of adatoms that governs the material transport on the
substrate surface towards an evolving island.
Here, we shall report about recent DFT calculations to investigate
this aspect for a particular example, the adsorption and surface
diffusion of indium during heteroepitaxy of InAs on GaAs(001).  This
system is characterized by a lattice mismatch $(a_{\rm InAs}-a_{\rm
GaAs})/a_{\rm GaAs} = \Delta a/a = 7.2$\%.  To be specific, we study
In adatom diffusion on the $c(4\times 4)$-reconstructed GaAs(001)
surface that is experimentally observed in the initial stages of InAs
deposition in the low-temperature growth regime, $T < 500^{\circ}$C
\cite{joyceASS:98}.  The geometric structure of this surface is
described as follows: On top of a complete As surface layer, the
$c(4\times 4)$ reconstruction has rows of As dimers running in the
$[\bar 1 1 0]$ direction, with units of three As dimers interrupted by
a dimer vacancy.  The geometric structure of the surface, together
with the PES for In diffusion obtained from DFT calculations, is shown
in Fig.~\ref{InGaAsPES}.  The calculations show that it is the dimer
vacancy site ${\mathbf{A}}_1$ where a deposited In atom is adsorbed
most strongly.  Diffusion is dominated by hopping through the
transition state ${\mathbf{T}}_1.$   
From an analysis of the PES we conclude that diffusion is activated by
0.65 eV and slightly anisotropic.

Since the elastic distortions in the vicinity of strained
heteroepitaxial islands vary slowly on the atomic scale, we can study
the effect of strain, $\varepsilon$, within the first-principles
approach by performing DFT calculations for uniformly strained slabs
that mimic the local strain state $\varepsilon({\mathbf{r}_{\|}})$ of
the substrate surface, $\mathbf{r}_{\|}$ being a coordinate parallel
to the surface.  In particular, we investigate the strain dependence
of the In binding energy, $E_b(\varepsilon)$.  This quantity is
crucial for the lateral variation of the In concentration on the
surface when a stationary average In coverage is maintained by an
equilibrium between supply from an atomic In beam and loss due to
evaporation of In. In this case, the adatom density is given by
\begin{equation}
  n({\mathbf{r}_{\|}})=
    n_0\exp[E_b(\varepsilon({\mathbf{r}_{\|}}))/k_{_{\rm{}B}} T].
\label{n_of_r}
\end{equation}
Since the nucleation probability is an increasing function of the
adatom density $n$, the surface regions with enhanced adatom binding
due to strain will show an enhanced probability of island nucleation.
The described effect of strain is most pertinent to QD stacks: The
vertical correlation of quantum dots in adjacent layers is controlled
by the nucleation of the first islands on the capping layer. The
strain fields of buried islands in deeper layers affects the adatom
density and thus the nucleation probability on the capping layer.  For
a single buried InAs island in a GaAs capping layer, calculations
using continuum elasticity theory have shown that the surface lattice
of the capping layer is expanded in the region above the island
\cite{holy:99}.  In addition, our DFT calculations demonstrate that
the binding of In adatoms is enhanced for positive (tensile) strain
and decreased for negative (compressive) strain. The effect can be
described approximately by a linear strain dependence of the binding
energy,
\begin{equation}
E_b(\varepsilon) = 2.2\,{\rm eV} + \varepsilon  \times 3.8\,{\rm eV}.
\label{Eb_of_eps}
\end{equation}
The linear variation of the binding energy with strain can be
rephrased in terms of a change of the surface stress induced by the
adsorbate.  Apart from the linear effect, the calculated data also
show a non-linear contribution to the strain dependence of binding
energies.
Given the elastic expansion of the capping layer above a buried
island, combining (\ref{n_of_r}) and (\ref{Eb_of_eps}) puts us in
position to conclude that the growth of QD stacks in the InAs/GaAs
materials system is characterized by preferred nucleation of new InAs
islands above the largest buried islands in the layer below. This
result explains the observed correlation of InAs dots between
subsequent layers in the GaAs matrix. Furthermore, it has been pointed
out that the dominance of large islands acting as a seed for the
nucleation in the new layer gradually improves the regularity of the
in-plane orientation of quantum dots as well as their size homogeneity
\cite{tersoff:96}.
    
Another possibility how strain could affect the kinetics of island
growth is by changing the mobility of adatoms diffusing on the
surface.  During the growth process, the growing islands compete for
the deposited material.  A self-limiting growth mechanism is
conceivable that leads to a narrow island size distribution, provided
that the flow of material towards the largest islands is hindered by
their surrounding strain field. For materials systems of interest for
quantum dots, such as Ge/Si and InAs/GaAs, the free-standing
heteroepitaxial islands are under compressive strain at their base,
while the substrate beneath the island is expanded.  As a consequence
of this expansion, the substrate surface {\em around} an island is
under {\em compressive} strain (see, e.g., Fig. 2 in
Ref.~\cite{moll:98}).

For islands on the $c(4 \times 4)$-reconstructed GaAs surface,
(\ref{Eb_of_eps}) implies that the binding of a diffusing In adatom
is reduced for sites close to the island. In other words, the island
gives rise to a repulsive potential that raises the bottom of the well
of the migration potential of an adatom.  We have also investigated
the strain dependence of the energy of the transition state
${\mathbf{T}}_1$ with the help of DFT calculations.  For diffusion
near the island, the transition states are also raised in energy,
albeit not by exactly the same amount as the binding state.
In summary, our calculations show, for the particular case of
coherently strained InAs islands surrounded by the $c(4 \times
4)$-reconstructed GaAs surface, that the strain field around an island
will hinder the diffusive flow of material towards the island and thus
have a self-limiting effect on island growth. In most experimental
situations, the InAs islands grow in the presence of a wetting layer
on the GaAs substrate.  The effect of strain on the diffusion on top
of a wetting layer is subject to ongoing research.

\begin{figure}[tb]
\epsfxsize=\columnwidth
\centering
\epsffile{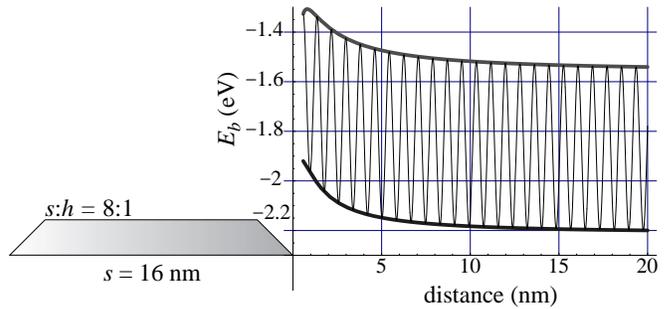}
\vskip0.5cm
\caption{Migration potential (oscillating curve) for an In adatom
approaching perpendicular to a very long, coherently strained InAs
island on the $c(4 \times 4)$-reconstructed GaAs(001) surface.  In
addition to the diffusion potential due to the atomic structure of the
surface, the strain field in the substrate induced by the island gives
rise to a repulsive potential that lifts both the binding energies
(thick lower line) and transition state energies (thick upper line)
close to the island.}
\label{migration}
\end{figure}

A scenario can be worked out explicitly for a simple geometry, a very
elongated, shallow island that is coherently strained
(Fig.~\ref{migration}).  Within the framework of continuum elasticity
theory, the strained island gives rise to a change of the stress
component parallel to the surface across the island edge, denoted by
$\Delta \sigma$.  The resulting force monopol generates a strain field
in front the island that can be calculated analytically within linear
elasticity theory (see Ref.~\cite{tersoff:93} and references therein).
If $x$ measures the distance for perpendicular approach to a
quasi-one-dimensional island with its edge at $x=0$, the strain field
has the form
\begin{equation}
  \varepsilon (x) = \eta\, \Delta \sigma \, \tan \theta \, 
                    \ln\!\left| \frac{P_2(x)}{Q_2(x)}\right|.
\label{1dstrain}
\end{equation}
The island geometry (height $h,$ base length $s,$ and the tilt angle
of the side facets $\theta$) determines the coefficients in the
second-order polynomials $P_2(x)$ and $Q_2(x),$ whereas the elastic
properties of the materials system enter the prefactor $\eta.$ An
explicit calculation for an elastically isotropic substrate yields
$P_2(x) = (h \cot \theta + x)(s - h \cot \theta +x)$, $Q_2(x) =
x(s+x)$. The prefactor is given by $\eta = 2(1+\nu)(1-\nu)/(\pi Y)$,
where $\nu$ is the Poisson ratio, and $Y$ is Young's modulus of the
substrate.  Neglecting the effect of elastic relaxation of the island,
the magnitude of $\Delta \sigma$ is estimated to be $\Delta \sigma
\approx Y \Delta a/a$, i.e. it is proportional to the lattice mismatch
$\Delta a/a$ between the island and the substrate.
By combining the DFT result expressed in (\ref{Eb_of_eps}) with the
knowledge of the long-range strain field obtained from linear
elasticity theory, we can assess the effect of strain on the transport
of deposited material towards an island.  Figure~\ref{migration}
illustrates a result for a particular geometry obtained by
substituting $\varepsilon(x)$ from (\ref{1dstrain}) into
(\ref{Eb_of_eps}), for an InAs island on GaAs with height $h=2$ nm
and base length $s=16$ nm.  As can be seen from the figure, the effect
of strain leads to a repulsive potential with a strength of up to
$0.2$ eV, that affects both the binding energy and, to a slightly
smaller extent, the diffusion barriers for an In adatom that attempts
to approach this island.  This repulsive interaction can significantly
slow down the speed of growth of strained islands.  It is worthwhile
to perform further investigations of the growth kinetics to study its
effect in detail.

Thus our investigations provide an example for a materials system (In
diffusion on strained GaAs(001)-$c(4 \times 4)$) with the unusual
property that smaller islands grow faster than larger islands: Since
the strain field around an island becomes stronger with increasing
island size, the effect of strain leads to retarded growth of larger
islands, and gives the smaller islands the chance to catch up during
growth.  For a simple example, where two very elongated islands
compete for the flux of In atoms deposited between them, we find that
strain-controlled diffusion indeed tends to equalize the size of the
two islands while they are growing \cite{Penev:01}. This effect has 
been proposed as one of the factors responsible for the narrowing of the 
island size distribution that is desirable from the point of view of
the applications~\cite{madhukar:96,kudovely:99}.

\section{Conclusion}
\label{Conclusion}

We have presented investigations of the kinetical aspects of epitaxy
using a combination of methods from statistical mechanics and
electronic-structure theory in order to treat phenomena at surfaces
such as adsorption, diffusion, island nucleation and growth.  The
density-functional calculations used to obtain the energetics of these
processes constitute a parameter-free approach that enables us to make
predictions independent from experimental input.  Following this
approach, computational modeling has evolved into an independent
research tool complimentary to experiments, whose outcome can and
should be tested against experimental data.  The use of multi-scale
modeling techniques, such as the combination of electronic structure
calculations with kinetic Monte Carlo simulations of kinetics, makes
it possible to retain the attractive features of first-principles
calculations, while enabling us at the same time to address problems
in physics or chemistry of surfaces on the experimentally relevant
length and time scales.  The combination of experimental research and
theoretical concepts developed from computer simulations eventually
results in an improved understanding of the processes behind epitaxial
growth on the atomic level.  Moreover, computational physics enables
us to address subtle aspects of the kinetics of growth that are hard
to probe directly in an experiment, but can have important
consequences for the growth morphology. As an example, we have
discussed the role of strain for the material transport in
heteroepitaxial systems.

\section{Acknowledgments}
\label{Acknowledgments}

We would like to acknowledge fruitful discussions with F.~Grosse,
M.~Itoh, T.~Mishonov, C.~M.~Morgan, and D.~D.~Vvedensky.  This work
was supported by the Deutsche Forschungsgemeinschaft, Sfb 296.

\end{document}